\documentclass[mathleft
]{an}
\usepackage{graphicx}
\usepackage{times}
\begin{document}

\Pagespan{789}{}
\Yearpublication{2014}%
\Yearsubmission{2014}%
\Month{11}%
\Volume{}%
\Issue{}%

\title{Companions around the nearest luminous galaxies: segregation and selection effects}
\author{I.D. Karachentsev\inst{1,2}\fnmsep\thanks{Corresponding author:
  \email{ikar@sao.ru}\newline}
\and  Y.N. Kudrya\inst{3}
}
\titlerunning{Companions around the nearest luminous galaxies}
\authorrunning{I.D. Karachentsev \& Y.N. Kudrya}
\institute{Special  Astrophysical  Observatory,  Russian  Academy  of  Sciences, Russia
\and Leibniz-Institut f\"{u}r Astrophysik, Potsdam, Germany
\and Taras Shevchenko National University of Kyiv, Ukraine}

\received{17 November, 2014}
\accepted{}
\publonline{later}


\abstract{
Using the ``Updated Nearby Galaxy Catalog'', we  consider different properties
of companion galaxies around luminous hosts in the
Local Volume. The data on stellar masses, linear diameters,
surface brightnesses, HI-richness, specific star formation 
rate ($sSFR$), and morphological types are  discussed for members 
of the nearest groups, including the Milky Way and  M~31 groups,  as a
function of their separation from the hosts. 
Companion galaxies in  groups tend to have lower stellar
masses, smaller linear diameters and fainter mean surface brightnesses
as the distance to their host decreases. The hydrogen-to-stellar 
mass ratio of the companions increases with their linear
projected separation from the dominant luminous galaxy. This tendency 
is more expressed around the bulge-dominated hosts. While linear
separation of the companions decreases, their mean $sSFR$ becomes
lower, accompanied with the increasing  $sSFR$ scatter. Typical
linear projected separation of dSphs around the bulge-dominated 
hosts, 350 kpc, is substantially larger than that around the disk-dominated
ones, 130 kpc. This difference probably indicates the presence of
larger hot/warm gas haloes around the early-type host galaxies.
The mean fraction of dSph (quenched) companions in 11 the nearest 
groups as a function of their projected separation $R_p$ can be expressed  
as $f(E) = 0.55 - 0.69 \times R_p$. The fraction of dSphs around the Milky
Way and M~31  looks much higher than in other nearby groups because the 
quenching efficiency dramatically increases  towards the ultra-low mass 
companions. We emphasize that the observed properties of the Local 
Group are not typical for other groups in the Local Volume due to the role
of selection effects caused by our location inside the Local Group. }

\maketitle

\section{Introduction}
The classical study of Dressler (1980) presented numerous observational evidence on the
segregation of galaxies by morphological types: the higher the number density of galaxies, the
 higher among them is the  fraction  of early-type objects with  old stellar populations, with low-gas abundances and slow 
star formation rates. Modern mass surveys of galaxies in different color bands like the Sloan 
Digital Sky Survey (Abazajian et al. 2009) have confirmed the fact, which has already become trivial  that many 
global properties of galaxies strongly depend  on the density of the environment in which 
the studied galaxy dwells. 

There are many publications, considering the effects of segregation of galaxies in  groups and clusters along 
their radius. There is also extensive literature which offers a variety of mechanisms to explain the observed 
morphological segregation: successive merging of galaxies due to the dynamical friction, sweeping out
of gas from dwarf galaxies as they pass through the dense haloes of massive neighbors or through the common 
hot gaseous medium in clusters and groups. 

There have been numerous studies in recent years  examining the effects of segregation based on the data  of
 N-body simulations  (Guo et al. 2001, Hirschmann et al. 2014, Wheeler et al. 2014, Slater \& Bell 2014).
It is well-known that more than a half of  galaxies are united in groups and clusters. 
A larger part of them  belongs to the population of small groups, such as our  Local Group. 
This is why the results of numerical simulations are 
usually compared with the observational characteristics of the Local Group (Libeskind et al. 2010, 2013, Knebe et al. 2011),
 considering its properties typical for the groups of galaxies in the present epoch (z = 0). 
However,  by a number of its average parameters the Local Group significantly 
differs from the other neighboring groups (Karachentsev \& Kudrya 2014, Karachentsev et al. 2013a). One of the major reasons 
of these differences is the observational selection effect conditioned by the location of the observer within the Local Group.
 Entwinement of the intrinsic segregations with the  observational selection effect introduces distortions 
that must be considered when comparing the results of N-body simulations with the observational data. 
In this paper we try to separate the effects of segregation and selection, considering  different properties of the
Local Group among the number of observed characteristics of a dozen other neighboring groups, the population and 
structure of which have been studied now with  appropriate thoroughness.

\section{The sample of nearby groups}
The most complete sample of nearby galaxies, presented in the ``Updated Nearby Galaxy Catalog'' (= UNGC, Karachentsev et al.
 2013a)  contains the data on distances, $D$, stellar masses, $M^*$,  hydrogen masses, $M_{HI}$, morphological
 types (de Vaucouleurs et al., 1991), $T$, and  star formation rates, $SFR$, for $\sim800$ galaxies located in a sphere of 
  11 Mpc radius. In  about half of these galaxies, the distances are measured by the Hubble Space Telescope
with the accuracy of at least 10\%. For each  UNGC-galaxy the  ``tidal index''  $\Theta_1$ has been determined as 
follows:

\begin{equation}
\Theta_1= \max[\log(M^*_n/D^3_n)] + C, \,\,\,\, n=1,2,...\, N, 
 \end{equation}
 which distinguishes among the plenty of nearby galaxies the most significant neighbor, whose tidal force, 
 $F_n\sim M^*_n/D^3_n$ dominates all other neighbors. Galaxies with one common gravitationally dominant neighbor, called 
 the ``Main Disturber'' = MD form a kind of a family or suite of this MD.  Both
physical companions of a given MD as well as distant galaxies of the  general `field'   occur in this suite.  The value of constant $C=-10.96$ was chosen so that $\Theta_1=0$ when the Keplerian cyclic period of the galaxy
with respect to its MD equals the cosmic Hubble time, $1/H$. In this sense, galaxies with $\Theta_1 < 0$
may be considered as undisturbed (isolated) objects.  At the same time, the set of companions with  $\Theta_1 >0$ 
is quite consistent with the notion of a group of galaxies around the dominant MD. Here, calculating the constant $C=-10.96$,
we adopted that the total mass of each galaxy is 6 times its total luminosity in the K- band. Under such definition, 
the most remote members of the Local Group: Leo A, Tucana, DDO 210, and WLM have their $\Theta_1$ just around 0, being 
situated near the `zero velocity sphere' of the Local Group. It should be stressed that with this definition of a group 
of galaxies, no restrictions (usually subjective) were set  on the difference of  radial velocities and/or  mutual 
separations of the group  members (MD suite).

The principle of identification of the companions by the zones of gravitational dominance of 
their MDs made it possible to find in the Local Volume
a lot of suites with   $n=53$ to $n=1$  members. Their summary  is presented in Table 1 (Karachentsev et al.
2014), a machine-readable version of which is available at
 http://lv.sao.ru/lvgdb/article/suites\_dw\_Table1.txt.
 The kinematic and dynamic properties of 15 most populous groups (suites) were discussed in detail  by
Karachentsev \& Kudrya (2014). In general, the sizes, integrated luminosities and virial  masses of
 these suites correspond well to the current ideas about the typical parameters of groups of galaxies.

\section{The segregation of suite members by stellar masses, diameters and surface brightness.}

 At first, we had to exclude two groups from  15 richest suites (groups) of the Local Volume: one around IC~342  and another 
around NGC~6946,
since both of them are located at low galactic latitudes, where it is difficult to identify dwarf spheroidal  galaxies
(dSph), possessing  low surface brightness. The loss of statistics here was compensated by the necessary homogeneity of the sample.
 The remaining suites were divided into two categories: `Early-type MD' and `Late-type MD' according to the morphological
type of their principal galaxy. To the first one we attributed  groups with the MD type $T\leq2$, where the main galaxy is
dominated by a bulge. These are 5 groups (suits) around: NGC~3115 ($T=-1$),NGC~3368 ($T=2)$, NGC~4594 ($T=1$), NGC~4736 
($T=2$) and NGC~5128 ($T=-2$). To the second category we have attributed 6 suites with the MD type $T > 2$, where the host galaxy 
is dominated by a disk population. These are groups around: NGC~253 ($T=5$), M~81 ($T=3$), NGC~3627 ($T=4$), NGC~4258 ($T=4$),
NGC~5236 ($T=5$) and M~101 ($T=6$). Two principal galaxies in the Local Group: the Milky Way ($T=4$) and M~31 ($T=3$) are also
the disk-dominated MDs. However, we distinguished them in a separate third subsample, `LG', because the conditions of companion
search around the Milky Way and M~31 were significantly different from those at work in the remaining
 groups of the Local Volume.

The distribution of the companions by the projected separation $R_p$ from the main galaxy and logarithm of the stellar
 mass is represented in
three panels of Fig. 1. The stellar mass of galaxies was determined by their $K$-band luminosity, assuming the 
mass-to-luminosity ratio   $M^*/L_K=1\times M_{\odot}/L_{\odot}$  (Bell et al. 2003). Physical companions of the
main galaxies with the tidal indices   $\Theta_1\geq 0$ are shown by solid circles. 
Due to the errors in distance determination, reaching  $\sim(1-2)$ Mpc at the outskirts of the Local Volume, the membership
 of some galaxies in groups is fragile. Therefore, we included in the consideration  the likely  companions of the main galaxies
  having slightly negative
tidal indices of   $\Theta_1=(0, -0.5)$.  Some of them can be real members of the groups, and the other may prove to be the
 general field galaxies.
 These objects are shown in the figure by empty squares. The upper panel of the figure corresponds to the groups,   dominated 
by an early-type galaxy. The middle panel combines data for the groups with the main galaxies  of   late morphological types, and
the bottom panel shows the companions of MW and M~31. Two lines with a short and long stroke show the linear regression 
\begin{equation}
 \log M^* = a\times R_p+b 
 \end{equation}
for the companions with $\Theta_1\geq 0$ and $\Theta_1\geq -0.5$, respectively.

The main characteristics of these groups are presented in Table 1. The upper rows of the table contain the following information:
(1)  the average value of logarithm of  stellar mass of the group's main  galaxy with the error in  mean, 
(2)  the total number  of
companions with   $\Theta_1\geq-0.5$  in each category of groups, 
(3)   the average projected  separation of the companions, 
(4)   the mean logarithm of stellar mass of companions, 
(5)   the slope of the regression line, and its   error in the case of $\Theta_1\geq-0.5$,
(6)   the Fisher statistics parameter characterizing the significance of the regression slope.

 The following  conclusions can be drawn from these data.
 
\begin{itemize}	
\item The difference in stellar masses of the main galaxies of Early-type and Late-type groups is not significant, 
but the masses of MW and M~31 are notably inferior to the mass of the main galaxy in the other 11 groups.

 \item  There is no clear decrease in the mean stellar mass of companions from periphery to the center of the Early-type
 and Late-type groups, and only a marginally significant trend at a $2\sigma$ level is seen for the companions of the
  MW and M~31.
Apparently, its origin is due to a selection effect, since the ultra-faint dwarfs with $\log(M^*/M_{\odot})<6$ were beyond
 the  detection limit  at large distances from the MW (Belokurov et al. 2006, Willman et al. 2009), and the  
 zone of search for the 
low-mass companions around M~31 did not cover the far periphery of this group (Ibata et al. 2007, Martin et al. 2009).

\item Inclusion or disregard of the companions with  $\Theta_1=(0,-0.5)$,  which can conditionally be called ``objects 
of the first infall towards the group center''   do not significantly affect the slope of the regression lines.

\end{itemize}

Figure 2 shows the distribution  of the companions by their Holmberg diameter  $A_{26}$
 and projected separation from the
main galaxy for  the E-groups (the top panel),  L-groups (middle one) and the Local Group (lower panel). 
The symbols in this figure are the same as in the previous one. The average value of  logarithm of the diameter, the slope of the regression line   $(a \pm \sigma$) and the significance parameter  of the slope by Fisher are presented, respectively, 
in rows (7--9) of Table 1. It can be concluded from the data of Fig. 2 and Table 1 that  the average linear
diameter of the companions decreases from periphery to the center of the group, showing the trend on  
$a/\sigma\sim(2 - 3)$ level. A similar effect is noticeable both for the companions of the MW and M~31,  
although it may partly be due to the above-noted selection effect. A possible physical cause of the observed trend in 
linear size of dwarf companions suggests a tidal stripping their outer parts, occurring in the tight vicinity of massive 
host galaxies (Mateo 1998, Nichols et al. 2014).

Another important parameter of galaxies is the average  surface brightness (SB). In the UNGC-catalog it is determined within the
Holmberg isophote  26.5 mag/sq.sec in the B-band. The distribution of the companions by their average surface brightness
is shown in  three panels of Fig. 3, where all the symbols are the same as in the previous figures. The mean SB of the 
ensemble of companions around the Early-type and Late-type MDs and around MW+M~31 are shown in  line (10) of Table 1. The last 
two rows indicate the value of the regression slope, $a\pm \sigma$, and its statistical  Fisher parameter. As one can  see from
these data, the surface brightness of the companions  shows a brightening tendency with an increase in their projected separation.
However, there is no significant trend in the E-groups, it is only pronounced at a $a/\sigma\sim2$ level in the companions of 
L-groups and the LG. One reservation should be done here. In some nearby ultra-faint dwarfs, resolved into stars, their central
SB often lies below the Holmberg isophote. In such cases, the average SB, presented in UNGC catalog, was estimated 
within the effective diameter of the dwarf, containing a half of its integral luminosity. This fact brings uncertainty in the
physical interpretation of the data. Anyway, as followed from three panels of Figure 2, the most of extremely low surface
brightness dwarfs, having the average SB fainter than  26.5 mag/sq.sec, occure in the neighbourhood of the Milky Way and M~31.

\section{HI-richness in population of the nearby groups.}

As it was shown by Haynes \& Giovanelli (1984), the ratio of the neutral hydrogen mass in the galaxy to its stellar mass,
 $M_{HI}/M^*$,\  is significantly smaller  in the central regions of clusters, as compared with the  population of the general field.
Subsequently, this  trend of HI-richness along a radius was investigated by many authors in systems of varying degree of density 
and population (Cortese et al. 2008, Taylor et al. 2012, Nichols \& Brand-Hawthorh 2013, Phillips et al. 2014).

The distribution of galaxies in nearby groups by ratio $M_{HI}/M^*$ and  projected separation from the main galaxy is shown in
three panels of Fig. 4. The upper and middle panels correspond to the E- and L-groups, the lower panel reproduces the data for  companions of the MW and M~31. The designations of galaxies are the same as in  previous figures. 
Additionally, crosses mark the companions in which only an upper limits of  HI-flux was measured. A long-stroke straight line  is drawn in view of the upper limits of the HI-flux
in contrast to Figures 1-3, where the line is drawn through the points with $\Theta_1\leq -0.5$.

The data on these groups are presented in Table 2. The left columns of the data correspond to the galaxies in groups with
 measured HI-masses, while the right columns also include the galaxies, where  only the upper limits of  hydrogen masses are known. The Table 2 rows contain:
(1)  the total number of companions in groups of each category, (2)  the average projected separation of the companions,
 (3)  the average
difference of  logarithms of hydrogen and stellar mass, (4)  the slope of the regression line  $\log (M_{HI}/M^*)= a\times R_p+b$, where
$R_p$ is expressed in Mpc, (5)  the statistical Fisher significance of the  ``$a$'' parameter.

We can make the following conclusions from Figure 4 and Table 2.

\begin{itemize}

\item The behavior of the average    $M_{HI}/M^*$ ratio  with  separation from the main galaxy of the group shows the expected 
 deficit of HI-richness in  central parts of the group. The average  effect reaches a factor of $\sim(3-5)$.

\item  The deficit of HI-richness is more pronounced in groups,  dominated by early-type galaxies. This may indicate the
existence around massive E, S0, Sa-galaxies of a more extended halo of warm/hot gas, with which the interaction
leads to the sweeping of neutral gas from the companions.

\item Account of galaxies with the estimates of the upper limit of $M_{HI}$ significantly enhances the observed effect.

\item The correlation of  $M_{HI}/M^*$ ratio  with the projected separation $R_p$ also manifests itself for the MW  
and M~31 companions. As the Local Group contains  much less massive companions with  $\langle\log M^*\rangle=6.3$ 
(see Table 1),
then the sweeping of gas from them should be more effective than that in the massive companions like LMC and SMC.
 
\end{itemize}

\section{Projected separation and star formation rate in the MD companions.}

One of the most reliable indicators that allow confident division of passive and active galaxies is their specific 
star formation rate,  $sSFR\equiv SFR/M^*$, expressed in   [yr$^{-1}$] units. 
  Here, the integral SFR of a  galaxy is usually estimated either by the 
integral flux in the emission  $H\alpha$ line, or by the flux in the far ultraviolet ($FUV$), measured by 
 GALEX satellite (Gil de Paz, 2007). Karachentsev \& Kaisina (2014) reviewed the data available in the UNGC catalog  on
 star formation rates in galaxies of the Local Volume depending on their tidal index  $\Theta_1$. 
 These data confirmed the known fact that the share of emission galaxies in the groups ($\Theta_1>0$) is significantly 
smaller than that of  field galaxies  ($\Theta_1<0$).   Here we 
found it useful to develop these data   not depending on   $\Theta_1$,  but on the linear projected separation of companion,
 $R_p$, 
with respect to the main galaxy of the group. Two lower panels of Fig. 5 represent the specific star formation rate of the
companions as a function of $R_p$ for 11 nearby groups and for the companions of MW and M~31. 
The star formation rate for them is determined  by  the  $H\alpha$ line flux. 
Two upper panels of Fig. 5 reproduce the same data calculated by the $FUV$ flux.
 Symbols in Fig. 5 are the same as in the previous figure. 

The average characteristics of the companions in groups of different categories are presented in Table 3. Left and right
 columns of the data correspond to the  
cases where the upper limits of the $H\alpha$  or $FUV$ fluxes of the companions were ignored, or have been taken up as
 the actual values. The structure of  Table 3 is similar to the previous ones. The  bottom part of Table 3 reproduces
  the sample of companions with
measured $H\alpha$ fluxes, and the  upper part  --- that with $FUV$ fluxes. We made a distinction between the $H\alpha$
 and $FUV$ samples, since the $SFR$ estimates based on them have some systematic differences discussed by Pflamm-Altenburg 
 et al. 2009, and Karachentsev \& Kaisina 2014.

The following conclusions can be made from the data of Fig. 5 and Table 3.

\begin{itemize}

\item The effect of reducing the average specific star formation rate with decreasing separation $R_p$ is clearly visible both in 
$H\alpha$ and  $FUV$ fluxes. The variation of $sSFR$ from the center to the periphery of the group reaches a factor of 10. The morphological type of the main galaxy has little effect on the amplitude of $sSFR$ trend.

\item A similar, but even more remarkable variation of $sSFR$ along the radius manifests itself in the companions of MW and M~31. 

\item For the entire ensemble of nearby groups, including the Local Group, the dispersion of  $sSFR$ in the companions increases from periphery to the group center. 
At the same time, as noted by Karachentsev \& Kaisina (2014) and Karachentsev et al. (2013),   individual
 $sSFR$ values do not exceed the upper limit of   $\max log[sSFR]=-9.4$ yr$^{-1}$. The presence of this limit, as well as its value itself are important characteristics of the process of star formation in the present epoch.
\end{itemize}

\section{ Segregation of companions by morphological types}

The above-mentioned effects of segregation of the companions by the relative abundance of hydrogen and specific star formation rate unavoidably manifest themselves in the form of morphological segregation effect, well known in the literature. Following de Vaucouleurs et al. (1991) 
we used, with minor modifications, such a relation between the numerical, $T$, and letter designation of types of galaxies: 
[--3, --2, --1]=[E, dSph], [0] = [S0], [1] = [S0a], [2] = [Sa], [3] = [Sab, Sb], [4] = [Sbc], [5] = [Sc], [6] = [Scd],
[7] = [Sd, Sdm], [8] = [Sm], [9] = [BCD, Im], [10] = [Ir], [11] = [HI cloud].

The distribution of morphological types of companions and the projected linear separation from the main galaxy (MD) is shown in Fig. 6. Its upper and middle panels correspond to the groups, where the MD  refers to the early 
($T\leq2)$ or late $(T> 2$) types. The bottom panel shows the distribution of companions around MW and M~31. 
Physical companions with  $\Theta_1 \geq 0$ are shown by solid  circles, and the probable companions (or the ``first infall'' 
objects) are marked by empty diamonds. The effect of morphological  segregation along the radius of  group is  quite 
clearly observed. 

Table 4 reproduces the mean values, $\langle R_p\rangle$ in Mpc, for the companions, divided into 4 categories according 
to their morphological types:  $T\leq 2, 3\leq T\leq 8, T=9$ and $T=10$.  
The numbers in parentheses indicate the number of galaxies
in each subsample. A comparison of the average  $\langle R_p\rangle$ for the physical companions with $\Theta_1\geq 0$  
(left column)  and companions with
   $\Theta_1\geq -0.5$ (right column) show how stable are the mean values, if we additionally include in 
   the sample probable remote companions as the ``first infall'' objects. Four twin columns of Table 4 correspond to division of  nearby groups according to the type of the main galaxy, which has been used above.

These data indicate that the average projected separation of the early-type companions is  2--3 times smaller than of the S-, BCD- and dIr-companions.
This segregation occurs both in the groups, dominated by galaxies with dominated bulges (E, S0, Sa), and in the groups, where MD refers to the late types. The same trend is visible for the members of the Local Group, in spite of the small statistics of the  late-type companions. Accounting for or ignoring of the possible companions with $\Theta_1=(0, -0.5)$  notably affects the value of  $\langle R_p\rangle$, especially for the late-type companions. 
In this sense, the $\langle R_p\rangle$  parameter is not quite a robust estimator of the
effect of morphological segregation in groups. (Note that the  trend of reduction of the average separation for the dIr-companions compared
with the   \{BCD+Im\} companions may indicate the presence among the companions with $T=10$ of some population of tidal 
 Holm-IX-type dwarf galaxies, which are located just beside the bright galaxies.)

Figure 7 shows the variation of the relative number of early-type companions in the nearby groups along the projected 
separation $R_p$.
Solid circles and triangles correspond to groups with the main galaxy of an early- or late- type, and the solid
squares mark the members of the Local Group. The intervals of the projected separation were chosen in order to
provide sufficient statistics in each of them. Only the assumed physical companions
with  $\Theta_1\geq 0$ were taken into account. From the analysis of these data we can draw the following conclusions.

\begin{itemize}

\item In general, the fraction of  early-type companions  around the main galaxies with   dominant bulges, 
 $f(E|E)=33/89=0.37$, does
not significantly differ from the relative number of  early-type companions around the  late-type main galaxies  dominated by disks,
 $f(E|L)=29/92=0.32.$  However, the characteristic separation of early-type companions in the  E-groups,  $\sim 350$ kpc, proves to be
significantly greater than in the L-groups,  $\sim 130$  kpc. This difference likely indicates the presence in the surrounding massive E, S0, and Sa-galaxies of more
extended warm gas haloes, which contribute to the transformation of star-forming  irregular dwarf companions
into the quenching spheroidal galaxies.

\item For the companions in all 11 nearby groups, the variation of the fraction   $f(E)$  along the radius is shown 
by empty diamonds indicating the statistical error. The  $f(E)$ vs. $R_p$  trend  looks pretty smooth. It can be represented 
by the linear regression
\begin{equation}
f(E)=0.55-0.69\times R_p 
\end{equation}
(shown by the thin solid line)
which goes to zero at $R_p$ = 0.80 Mpc. This is consistent with the statement of Geha et al. (2012) that dwarf dSph
galaxies with masses
$\log (M^*/M_{\odot})<9$ make up not more than 0.1\% among the field galaxies.

\item  For the  MW and M~31 companions,  the behavior of the     $f(E)$ vs. $R_p$  relation
is significantly different from the other nearby groups. 
Within the  $R_p=250$ kpc 
radius  the relative number of early-type companions in the LG exceeds 90\% instead of the average value of 40\% for the remaining nearby groups. This difference has an obvious explanation linked with observational selection: ultra-dwarf systems dominate among the companions of the MW and M~31.
They are easily exposed to the sweep-out of gas from their shallow potential wells.  
A special search for such ultra-dwarfs in the vicinity of 
M~81 (Chiboucas et al. 2009, 2013)
confirms that  they are dominated by quenching objects with no signs of star formation   (Kaisin \& Karachentsev 2013).

\end{itemize}

\section{Discussion and concluding remarks}

The distribution of companions around the Milky Way and M~31 by stellar mass, linear diameter, and surface brightness
 reveals   trends in the average values  of these parameters with the projected separation from the main galaxy. The trends of the  
$\langle M_*\rangle$,
  $\langle A_{26}\rangle$
and $\langle SB\rangle$ 
variables  of the same sign, but smaller amplitude can also be seen in the companions of 11 other neighboring groups. A comparison of the
effects of segregation in the Local Group with other nearby groups indicates a significant role of the observational selection effect
caused by our location inside the  Local Group. This is why the observed properties of the Local Group are not typical for the remaining groups of the Local Volume. This fact should be considered when comparing the results of N-body simulations with the observational data.
  
  In addition to the presence of ultra-faint dwarfs of extremely low surface brightness,
  resolved into stars only at nearby
distances, the Local Group stands out among the neighboring groups by the absence in it of blue compact dwarfs (BCD), as well as
HI-filaments and tidal dwarfs like  Holm-IX, BK3N, Garland, A0952+69, which are visible around M~81 (Makarova et al.
2002). It should be added here that the stellar masses of MW and M~31, their  virial masses, as well as the linear dimensions of the suite of companions
around the MW and M~31 are notably inferior to other high-luminosity galaxies in the Local Volume.

Both the Local Group and other nearby groups show  an increase towards  the center of the group a fraction
of  early-type companions with a reduced gas abundance per unit of stellar mass, as well as a  low star formation rate. 
The observed behavior of  $\langle M_{HI}/M^*\rangle$ and $\langle sSFR\rangle$ dependences on $R_p$
allows  to check various scenarios of dynamic evolution of the companions moving inside the haloes of massive main galaxies 
(Kormendy \& Freeman 2014, Yang et al. 2014). Based on these findings it can be assumed that the main early-type galaxies 
with dominant bulges possess more extended haloes composed of warm/hot gas than the disk-dominated galaxies of late types.

For the manifold of 11 most nearby and populated  groups of galaxies, the relative number of passive, quenching early-type
 companions decreases with the projected separation  as 
 $f_E(R_p)=0.55-0.69\times R_p$, where $R_p$ is expressed in Mpc. Formally,  
this implies that beyond  $R_p=0.8$ Mpc no dwarf spheroidal companions should be present. Indeed,
in the Local Group only the dwarf system Cetus with $\Theta_1=0.3$ and $D=0.78$ Mpc has $R_p=1.0$
Mpc with respect to M~31  as its host. However, in the Local Volume, there are two other examples 
of isolated dSph galaxies: KKR25 with $D=1.86$ Mpc and
     $\Theta_1=-1.0$ (Karachentsev et al. 2001, Makarov et al. 2012) and Apples~I with $D=8.3$ Mpc and $\Theta_1=-1.5$ 
     (Pasquali et al. 2005).  Recently, a third very isolated dSph, KKs3, at a distance of 2.23 Mpc was found
by Karachentsev et al. (2015).
   The search for other such  quenching orphan dwarfs is an extremely painstaking
observational  task that can be solved in terms of  deeper wide-field  sky surveys in the near future.

The assumed processes involving sweeping-out of gas  from the dwarf companions and locking of  star formation in them
while passing through the dense regions of  haloes of massive neighbors manifest themselves most effectively 
for the
companions with the smallest potential wells. Therefore,  the most nearby and most studied groups in the
Local Volume are the most suitable objects for the  analysis of  different segregation effects  in them.

 \acknowledgements
 We are grateful to the anonymous referee for comments that helped us improve the manuscript.
This work was supported by the grant of the Russian Foundation for Basic
Research   13--02--90407 Ukr-f-a, and the grant of the Ukraine F53.2/15.
IK acknowledge the support for proposal GO 13442
provided by NASA through grants from the Space Telescope Science Institute.

\clearpage

\begin{figure*}
\includegraphics[scale=0.4]{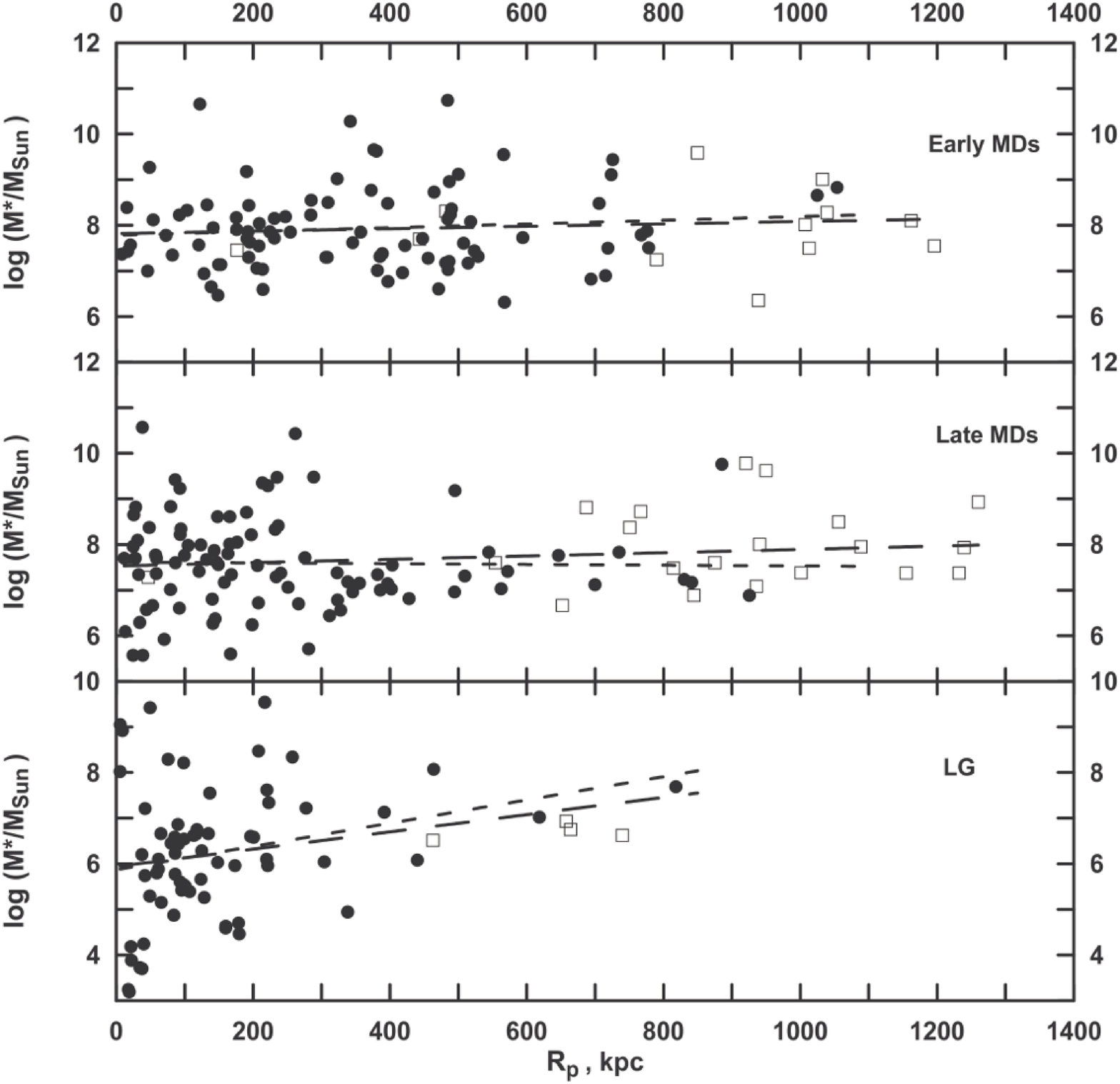}
\caption{The distribution of companions of nearby luminous galaxies by their stellar mass and projected separation. Solid circles represent
physical companions with the tidal index $\Theta_1\geq 0$, empty squares mark probable companions with   $\Theta_1=(0, -0.5)$. 
 The straight line with a short stroke corresponds to the regression for the physical  companions, the straight line with a  
 long stroke --- the one for the  combined sample. 
The top panel depicts the companions around 5 massive early-type galaxies:
NGC~3115, NGC~3368, 
NGC~4594, NGC~4736 and NGC~5128.
The middle panel  represents the companions around 6 massive late-type galaxies: NGC~253, M~81, NGC~3627, 
NGC~4258, NGC~5236 and M~101.
The bottom panel --- the companions of the Milky Way and M~31.}
\end{figure*}

\begin{figure*}
\includegraphics[scale=0.4]{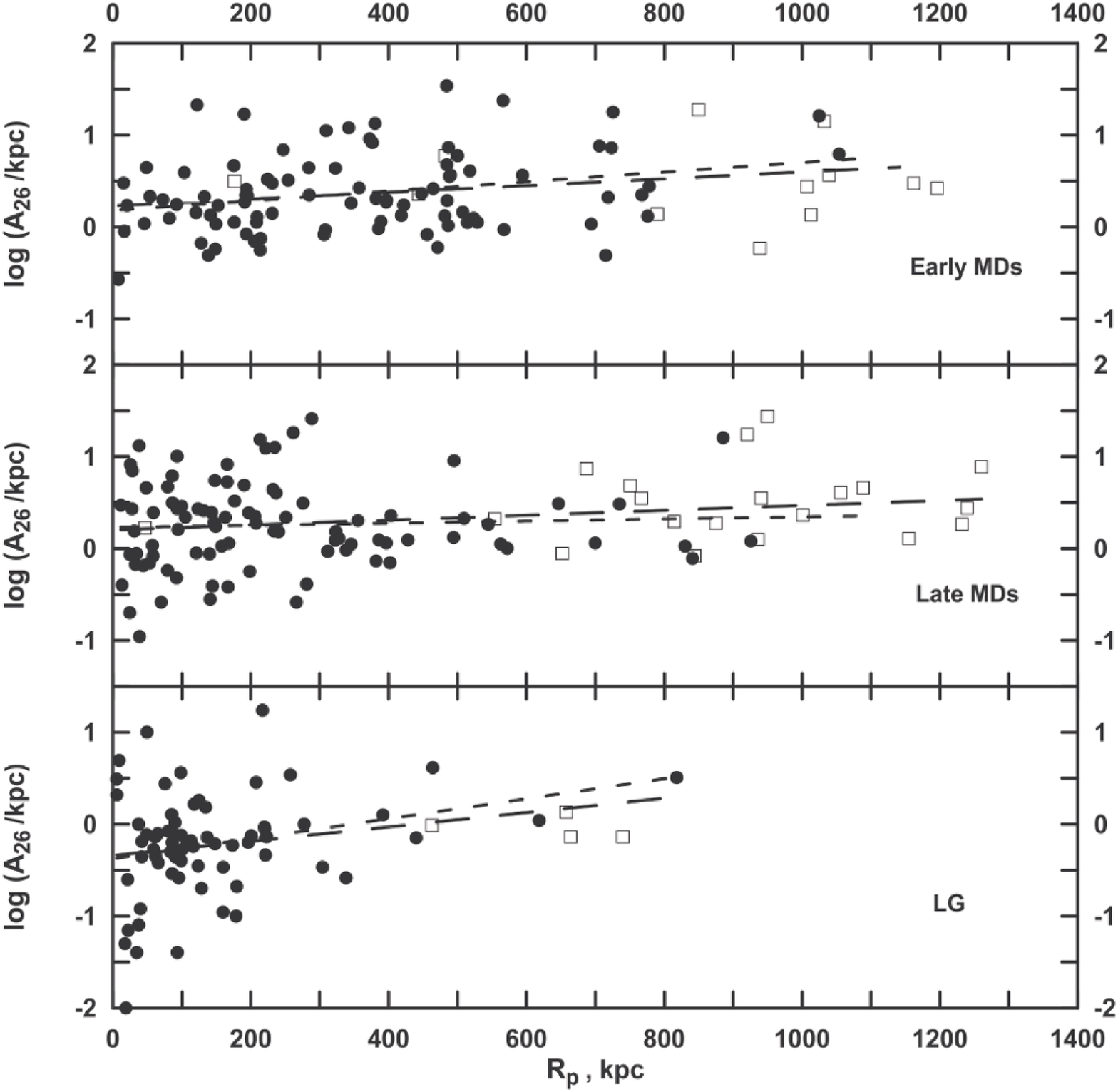}
\caption{The distribution of companions around nearby luminous galaxies by their linear Holmberg diameter and projected separation. 
The symbols are the same as in Fig. 1.}
\end{figure*}

\begin{figure*}
\includegraphics[scale=0.6]{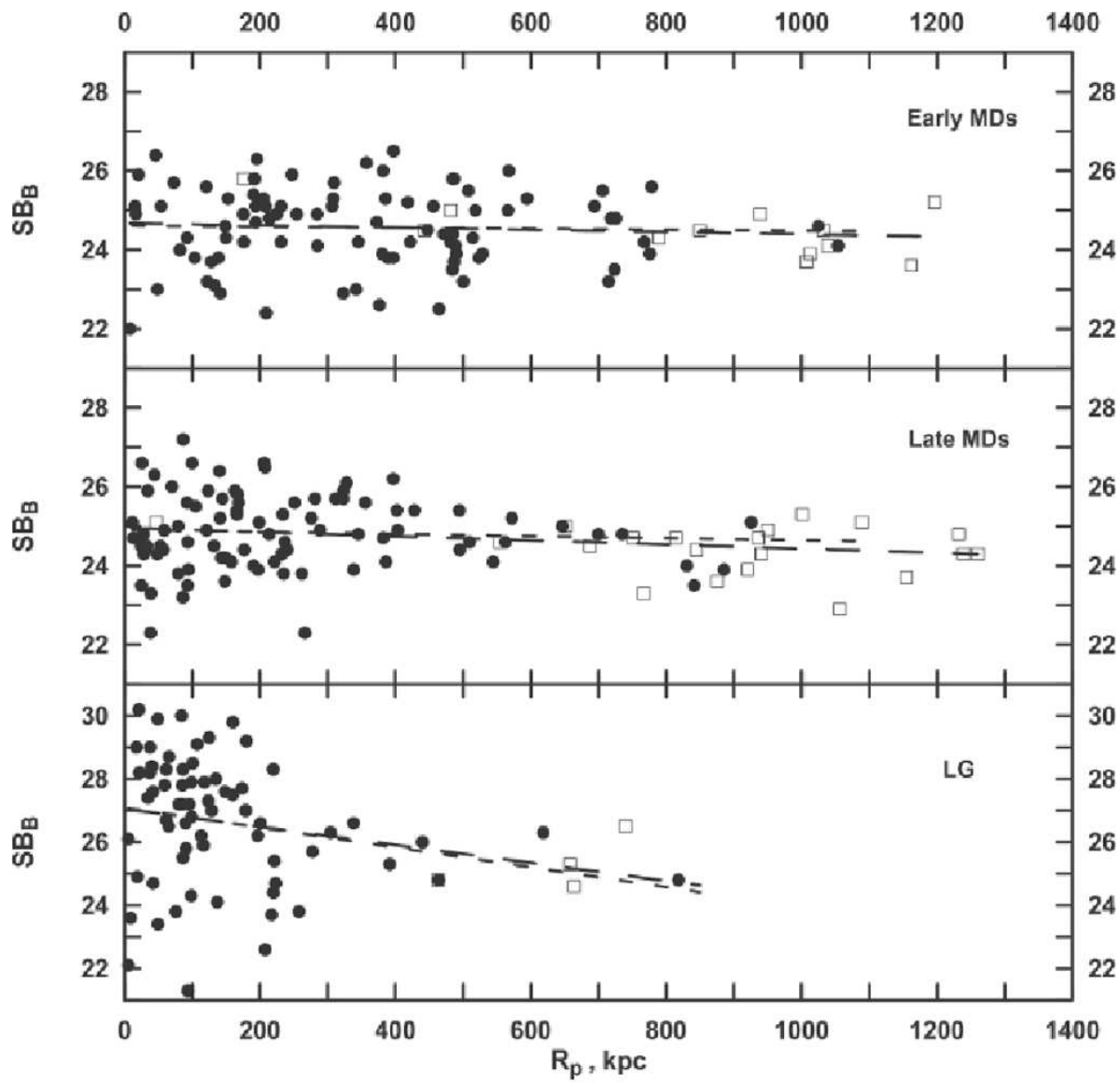}
\caption{The average surface brightnesses in the B-band and projected separation of companions around the nearby massive galaxies. 
The designations are the same as in the previous figures.}
\end{figure*}

\begin{figure*}
\includegraphics[scale=0.6]{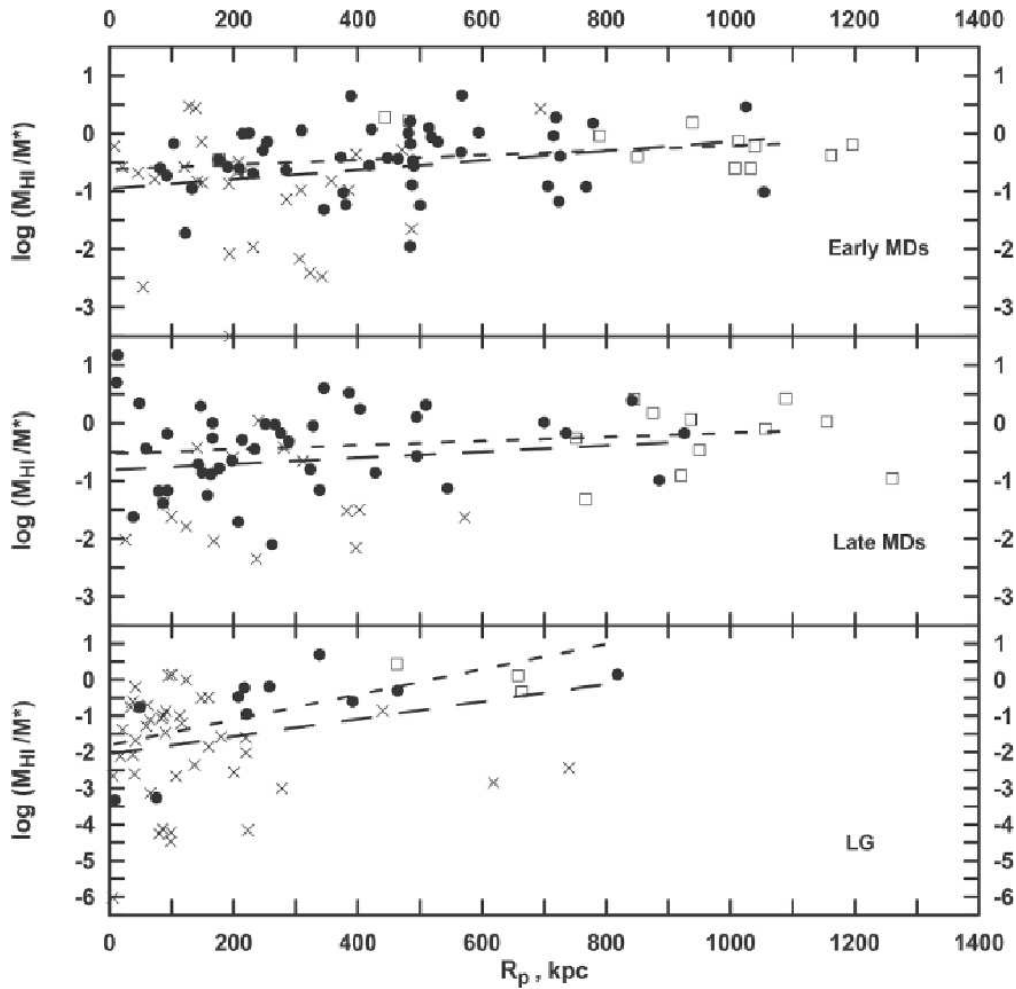}
\caption{The hydrogen-to-stellar mass ratio for companions of nearby bright galaxies at different 
 projected separation from  the main galaxy. The designations are the same as in the previous figures. 
 The crosses mark the companions, in which  only the top  limits of hydrogen masses are measured.}
\end{figure*}

\begin{figure*}
\includegraphics[scale=0.6]{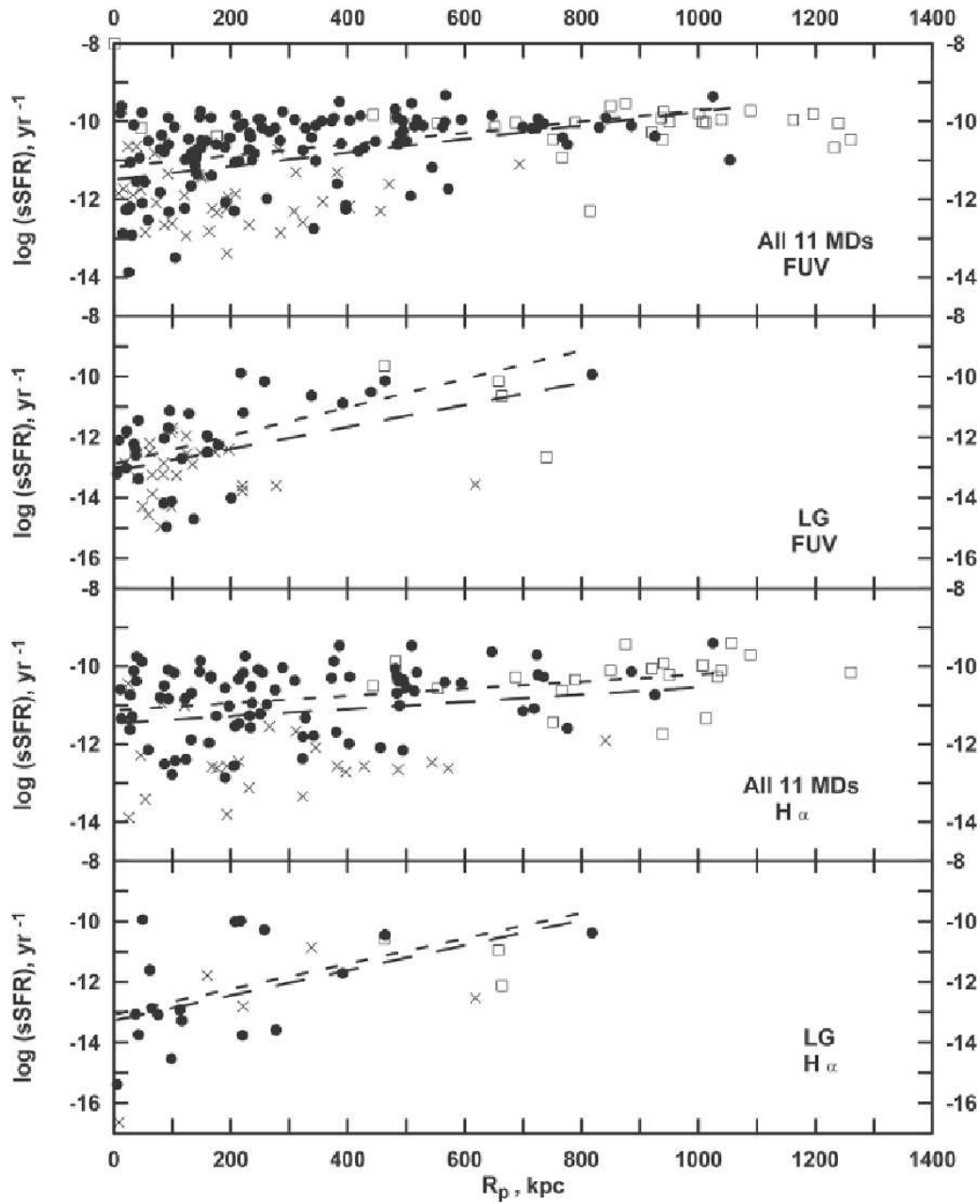}
\caption{Specific star formation rate vs. projected separation for the companions of nearby massive galaxies.
 The designations are the same 
as in the previous figure. The companions with  estimates of the upper limit of $SFR$  are indicated by crosses. 
Two lower panels represent the 
 $SFR$ estimates based on the   $H\alpha$ line flux for the companions around 11 luminous galaxies, and for the companions
of the MW and M~31. Two upper  panels correspond to the $SFR$ estimates by the $FUV$ fluxes measured with GALEX.}
\end{figure*}

\begin{figure*}
\includegraphics[scale=0.6]{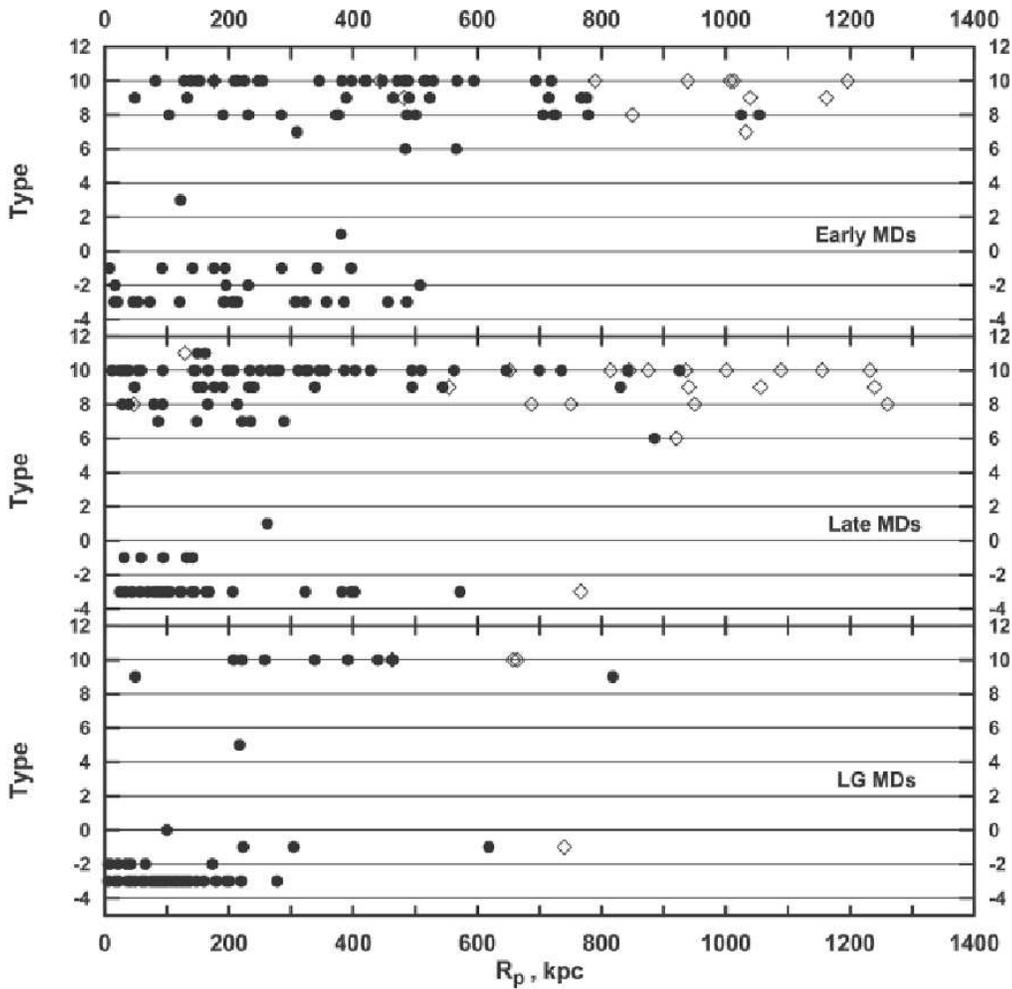}
\caption{Morphological type vs. projected separation for the companions around 5 luminous early-type galaxies 
(the top panel), around 6 luminous late-type ones  (the middle panel), and around the MW and M~31 (the bottom panel). 
Physical companions with   $\Theta_1\geq 0$ are shown by solid circles, and  probable companions having
  $\Theta=(0, -0.5)$ are shown by diamonds.}
\end{figure*}

 \begin{figure*}
\includegraphics[scale=0.6]{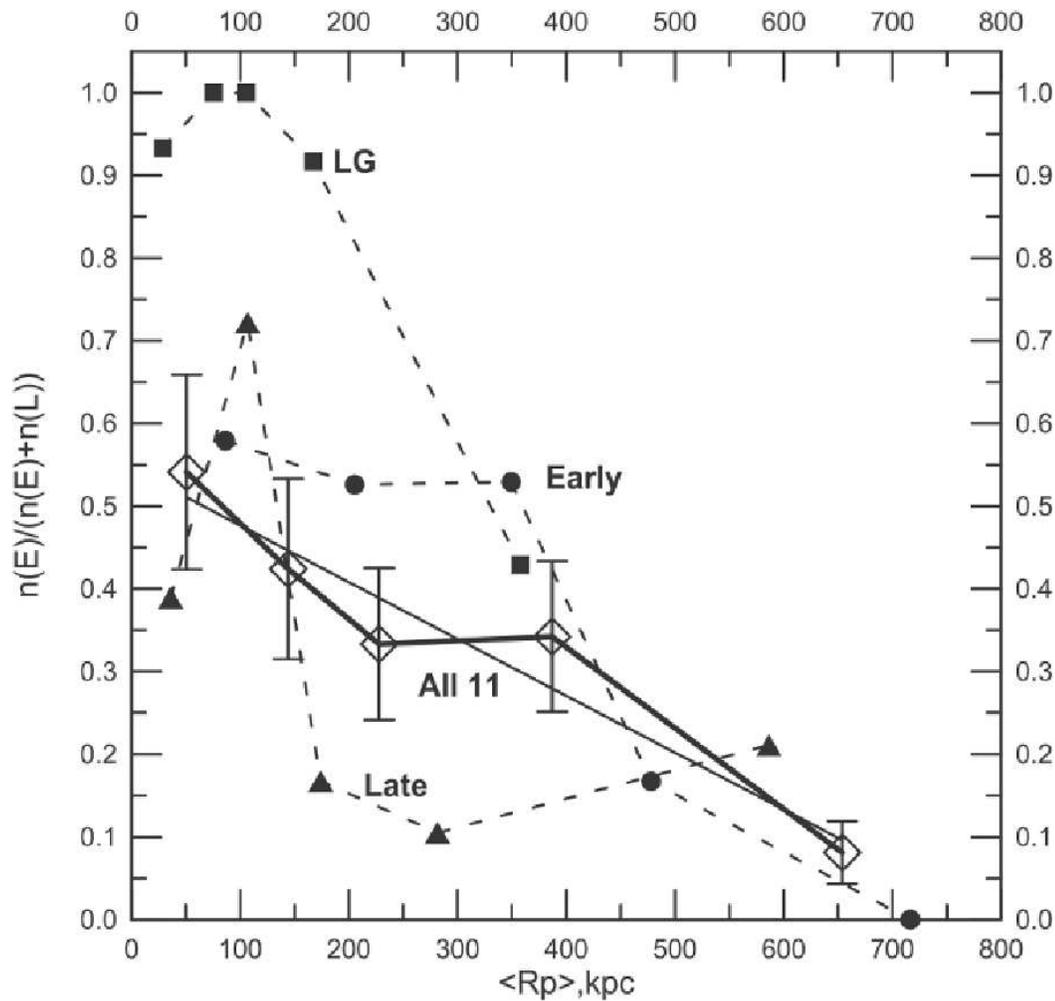}
\caption{The fraction of early-type companions at different projected separation around 5 massive
early-type galaxies  (circles), around 6 massive late-type galaxies (triangles) and around the MW and M~31 (squares). 
The data for  companions of all the eleven early+late-type galaxies are marked with open diamonds
 indicating statistical errors. The regression line (3) for them is shown  by  thin solid line.}
\end{figure*}

\clearpage

\begin{table*}
\caption{The parameters of  regressions  $y=a\times R_p+b$  for companions around the nearby luminous galaxies, including 
 probable companions with  $\Theta_1\geq -0.5$.}
\begin{tabular}{lcccc} \\ \hline

 Parameter   &   All 11 groups  & 
E--groups   &   
 L--groups   &  
LG \\ \hline
 $\langle\log M^*_{MD}\rangle$  &   10.93$\pm$0.07    &10.92$\pm$0.11 &  10.94$\pm$0.03  & 10.72$\pm$0.02\\
  Number     &       211        &    101     &      110      &     71\\
 $\langle R_p\rangle$, Mpc  &      0.38        &   0.41     &     0.36      &    0.17\\
 $\langle\log M^*\rangle$     &      7.79        &   7.93     &     7.66      &    6.27\\
 $a\pm\sigma$  &    0.36$\pm$0.21    & 0.26$\pm$0.31 &  0.36$\pm$0.29   & 1.89$\pm$0.91\\
  Fisher     &      2.84        &   0.71     &     1.59      &    4.29\\
\hline
 $\langle\log A_{26}\rangle$   &      0.34         &  0.38     &     0.30      &   --0.21\\
 $a\pm\sigma$  &    0.32$\pm$0.10     &0.37$\pm$0.15 &  0.27$\pm$0.13   & 0.78$\pm$0.36\\
  Fisher     &      10.7         &  6.37     &     4.14      &    4.76\\
\hline
   $\langle SB\rangle$      &     24.7          & 24.6     &     24.8     &     26.6\\
 $a\pm\sigma$   & -0.45$\pm$0.20   & --0.28$\pm$0.34  & --0.53$\pm$0.25  & --2.84$\pm$1.27\\
  Fisher      &     4.96          & 0.70     &     4.33     &     4.98\\ \hline
\end{tabular}
\end{table*}

\begin{table*}
\caption{The characteristics of HI-richness for companions around the nearby luminous galaxies. The right columns in each subgroup 
 correspond to the data taking into account the upper limit for  HI-fluxes.}
\begin{tabular}{p{1.0in}cc|cc|cc|cc} \\ \hline

 Parameter     & \multicolumn{2}{c}{All   11 groups} & 
   \multicolumn{2}{|c}{E-groups}   &
       \multicolumn{2}{|c|}{L-groups}   &       
       \multicolumn{2}{|c}{LG}\\
\hline
 Number         & 112  &  134   &    59 &   77    &  53   &  57    &  14  &   51\\
 $\langle R_p\rangle$, Mpc     & 0.48 &  0.33  &   0.53  & 0.36  & 0.44  & 0.29   & 0.35  & 0.16\\
$\langle\log(M_{HI}/M^*)\rangle $  &--0.39 & --0.66  &  --0.39 & --0.66  &--0.39&  --0.66  &--0.65 & --1.68\\
   $a$          & 0.29 &  0.68  &   0.31  & 0.81  & 0.28 &  0.53   & 3.09 &  2.41\\
 Fisher         & 2.69 &  5.07  &   1.75   &4.30  & 1.06 &  1.10   & 7.90 &  3.61\\
 \hline
 \end{tabular}
 \end{table*}

\begin{table*}
\caption{The characteristics of the specific star formation rate   (yr$^{-1}$)  for companions around the nearby luminous galaxies of
 different categories. The right columns correspond to the data taking into account the upper limit of $SFR$. The lower part
  of the table corresponds to the  $SFR$  estimates by the $H\alpha$ flux and the upper part --- to the 
 $SFR$ estimates by the $FUV$ flux.}
\begin{tabular}{lcc|cc|cc|cc} \\ \hline

 Parameter     & \multicolumn{2}{c}{All   11 groups} & 
   \multicolumn{2}{|c}{E-groups}   &
       \multicolumn{2}{|c|}{L-groups}   &       
       \multicolumn{2}{|c}{LG}\\
\hline
\hline
Number $(FUV)$      & 150  &   159  &    61   &  72    &   89  &   87   &    34   &  56\\
 $\langle R_p\rangle$, Mpc      & 0.42 &   0.29 &   0.47  & 0.34   &  0.38 &  0.25  &   0.22  & 0.15\\
$\langle\log(sSFR)\rangle$    &--10.60 & --10.99 & --10.50 &--11.05  & --10.67& --10.95 &  --11.94 &--12.57\\
   $a$             & 1.08 &   1.75 &   1.19  & 2.12   &  0.99 &  1.64  &   3.53  & 3.64\\
 Fisher          &28.5  &  29.5  &  16.0   &22.9    & 12.9  & 11.8   &  13.9   &12.1\\ \hline
Number($H\alpha$) &  102  &   105  &    39  &   40   &    63 &    65   &    21  &   23\\
 $\langle R_p\rangle$, Mpc     &  0.42 &   0.31 &   0.53 &  0.37  &   0.36&   0.27  &   0.25 &  0.21\\
$\langle\log(sSFR)\rangle$    &--10.74 & --11.19 & --10.56 &--11.08  & --10.85& --11.25  & --12.11 &--12.40\\
   $a$           & 0.91  &  0.92  &  0.25  & 1.95   &  1.12 &  0.22   &  3.39  & 4.15\\
 Fisher         & 13.4  &   4.09 &   0.44 &  6.74  &  11.3 &   1.35  &   5.75 &  5.80\\ \hline
 \end{tabular}
 \end{table*}
 
 \begin{table*}
 \caption{The average projected separation for the companions having different morphological types.}
 \begin{tabular}{l|cc|cc|cc|cc} \hline
 &\multicolumn{8}{c}{The mean projected separation (Mpc)}\\ \cline{2-9}
 \multicolumn{1}{c}{Companion type}&
  \multicolumn{2}{|c}{All 11 groups}&
   \multicolumn{2}{c}{E-groups}&
    \multicolumn{2}{c}{L-groups}&
     \multicolumn{2}{c}{LG} \\ \hline
 
 $T < 3$       &  0.19 &  0.20 &    0.22 &  0.22  &   0.16  & 0.18 &    0.11 &  0.12\\
               &  (61) &  (62) &    (32) &  (32)  &   (29)  & (30) &    (58) &  (59)\\
               &       &       &          &       &         &      &          &    \\        
$2 < T < 9$    & 0.38   &0.47   &  0.50  & 0.55   &  0.21   &0.39   &  0.22   &0.22\\
               &  (30)  & (38)  &   (18) &  (20)  &   (12)  & (18)  &   ( 1)  & ( 1)\\
                &       &       &          &       &         &      &          &    \\ 
 $T = 9$       &  0.38   &0.51   &  0.48  & 0.58   &  0.30  & 0.46   &  0.43   &0.43\\
               &  (21)   &(28)    & ( 9)  & (12)   &  (12)  & (16)   &  ( 2)   &( 2)\\
               &       &       &          &       &         &      &          &    \\
 $T = 10$      &  0.33   &0.44     &0.37  & 0.45    & 0.30  & 0.43    & 0.33   &0.41\\
               &  (67)   &(83)     &(30)  & (37)    & (37)  & (46)    & ( 7)   &(10)\\ \hline
\end{tabular} 
\end{table*}

\end{document}